\pdfoutput=1
\documentclass[apj]{emulateapj}
\usepackage[bookmarks=true,colorlinks=true,citecolor=blue,linkcolor=magenta,urlcolor=cyan]{hyperref}

\usepackage{natbib}
\usepackage{amsmath}
\usepackage{amssymb}
\usepackage{subfigure}
\usepackage{url}
\usepackage{CJK}

\renewcommand{\vec}[1]{ {\mathbf #1} }

\newcommand{\grad}{ {\bf \nabla } }

\newcommand{\Fig}{{Figure}}

\shorttitle{NLFFF Extrapolation of Filament Field}
\shortauthors{Jiang et al.}

\begin{document}
\begin{CJK*}{UTF8}{gbsn}

  \title{Nonlinear Force-Free Field Extrapolation of a Coronal
    Magnetic Flux Rope Supporting a Large-Scale Filament from
    Photospheric Vector Magnetogram}

\author{
  Chaowei Jiang \altaffilmark{1,2},
  S.~T. Wu \altaffilmark{1},
  Xueshang Feng \altaffilmark{2},
  Qiang Hu \altaffilmark{1}}

\altaffiltext{1}{Center for Space Plasma and Aeronomic Research, The
  University of Alabama in Huntsville, Huntsville, AL 35899, USA}

\altaffiltext{2}{SIGMA Weather Group, State Key Laboratory for Space
  Weather, Center for Space Science and Applied Research, Chinese
  Academy of Sciences, Beijing 100190}

\email{cwjiang@spaceweather.ac.cn, wus@uah.edu}
\email{fengx@spaceweather.ac.cn, qh0001@uah.edu}

\begin{abstract}
  Solar filament are commonly thought to be supported in magnetic
  dips, in particular, of magnetic flux ropes (FRs). In this Letter,
  from the observed photospheric vector magnetogram, we implement a
  nonlinear force-free field (NLFFF) extrapolation of a coronal
  magnetic FR that supports a large-scale intermediate filament
  between an active region and a weak polarity region. This result is
  the first in that current NLFFF extrapolations with presence of FRs
  are limited to relatively small-scale filaments that are close to
  sunspots and along main polarity inversion line (PIL) with strong
  transverse field and magnetic shear, and the existence of a FR is
  usually predictable. In contrast, the present filament lies along
  the weak-field region (photospheric field strength $\lesssim
  100$~G), where the PIL is very fragmented due to small parasitic
  polarities on both side of the PIL and the transverse field has a
  low value of signal-to-noise ratio. Thus it represents a far more
  difficult challenge to extrapolate a large-scale FR in such case. We
  demonstrate that our CESE--MHD--NLFFF code is competent for the
  challenge. The numerically reproduced magnetic dips of the
  extrapolated FR match observations of the filament and its barbs
  very well, which supports strongly the FR-dip model for
  filaments. The filament is stably sustained because the FR is weakly
  twisted and strongly confined by the overlying closed arcades.
\end{abstract}


\section{Introduction}
\label{sec:intro}

Filaments are thin
structures consisting of cool, dense plasma suspended in the tenuous
hot corona. They lie above polarity inversion lines (PILs) on the
photosphere, and are formed in filament channels where the
chromospheric fibrils are aligned with the PIL
\citep{Gaizauskas1997}.
Filaments can be found inside active regions (ARs, ``AR filaments''),
at the border of ARs (``intermediate filaments''),
and on the quiet Sun (``quiescent filaments'').
Most filaments eventually erupt and lead to coronal mass
ejections, which are major drivers of space weather \citep{Schmieder2012}.

The magnetic field plays a primary role in all the coronal processes
including filament formation because the plasma $\beta$ is low. It is
now commonly accepted that filament plasma is supported in magnetic
dips, in particular, within twisted flux ropes (FRs)
\citep[e.g.,][]{Rust1994SoPh, Chae2001ApJ,
  Ballegooijen2004}. Obtaining the three-dimensional (3D) magnetic
field that supports filaments is key to understanding their structure,
stability and eruption. Unfortunately, the 3D coronal field is very
difficult to measure directly. People thus seek numerical models to
construct the coronal field involved with filaments. The first
numerical model was developed by \citet{Aulanierfilament1} using
linear force-free field extrapolation from line-of-sight (LoS)
magnetogram, which is proven to be a powerful tool to simulate
filaments and related small structures like filament barbs
\citep{Aulanierfilament1, Aulanierfilament2, Aulanier2000,
  Aulanier2002, Dudik2008, Dudik2012}.  \citet{Ballegooijen2004} then
developed a nonlinear force-free field (NLFFF) model, the FR insertion
method, in which a FR with its axis following the targeted filament
channel is first inserted into a potential field environment and the
system is then relaxed to a force-free equilibrium. The FR insertion
method has been applied in modeling various filaments
\citep[e.g.,][]{Ballegooijen2004, Bobra2008, Su2011,
  Su2012,Savcheva2012}.


NLFFF extrapolation from the photospheric
vector magnetograms (VMs) is a more general method to reconstruct the
coronal field \citep{Sakurai1981, Wu1990, Amari1997,
  Wiegelmann2012solar}, regardless of the presence of filaments.
Unlike the FR insertion method, the NLFFF extrapolation can
reconstruct a FR naturally and in a straightforward way, if it
exists. Recent studies have reported many examples with FR
extrapolated from different VMs using various NLFFF codes
\citep[e.g.,][]{Yan2001, Regnier2004, Canou2010, Cheng2010, Guo2010,
Jing2010, Jiang2013NLFFF}. Some studies also show that the extrapolated FR
structure matches the related filament, e.g., good spatial correlation
of the FR dips with the filament channels \citep{Canou2010,Guo2010}.
We note that all these works are limited to relatively small-scale
filaments (length within tens of megameters) whose channel is close to sunspots and along the main
PIL with strong transverse field and magnetic shear. In such cases, the existence of a FR
is usually predictable by inspecting the VM. The sheared PIL even with
bald patches \footnote{Places on PIL where the transverse field is so
  strongly sheared that it points from the negative to the positive
  polarity, which is opposite to a potential field case
  \citep{Titov1999}.} usually indicates the presence of a coronal FR
\citep{Titov1999, Aulanier2010}.
On the contrary, many large-scale filaments (length up to hundreds of megameters),
like the intermediate and quiescent filaments, usually
lie above the photospheric weak-field region (field strength $\lesssim 100$G), where the
PIL is very fragmented due to small parasitic polarities on both side
and the transverse field is very noisy. As a result, the
signature of FR, e.g., unbroken PIL with strong magnetic shear, is
difficult to be observed directly from the magnetogram.
Thus in such conditions it is far more of a challenge to
extrapolate a large-scale FR from currently observed VM.

In this Letter, we report such an challenging NLFFF extrapolation which recovers a
large-scale FR that supports an intermediate filament with a length up to
300~Mm. The extrapolation is based solely on the SDO/HMI VM without
any other observation or artificial input. It reproduces a coronal FR matching
the filament recorded by AIA and H$\alpha$
strikingly well, which strongly supports the FR dip model for
filaments. We further study why the FR can keep its stability from eruption.


\begin{figure*}[htbp]
  \centering
  \includegraphics[width=0.8\textwidth]{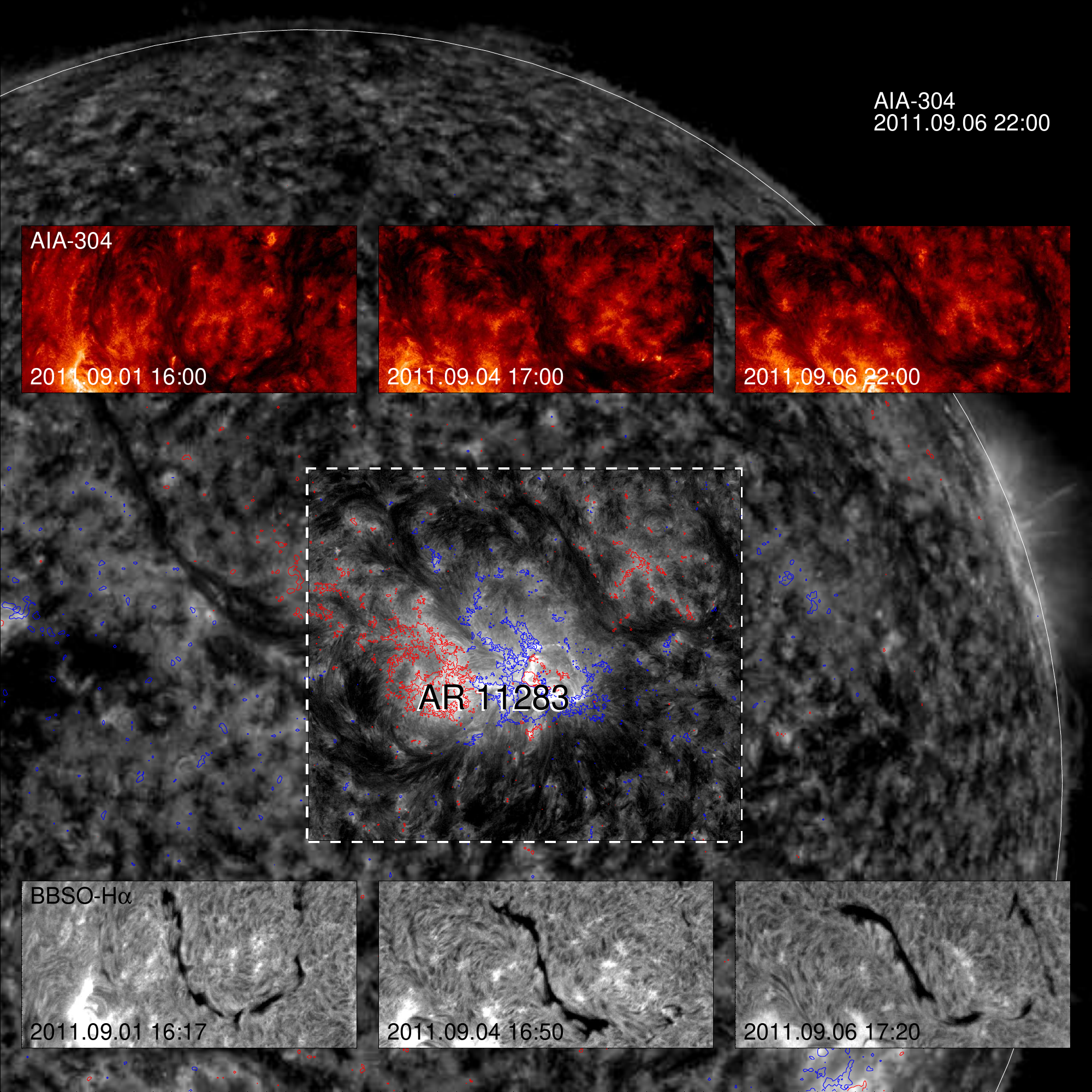}
  \caption{The background shows a large AIA-304 image at 22:00~UT on
    September 6, overlaid with contours of $\pm 100$~G (red/blue) for
    LoS photospheric magnetic field. The white arc is the solar
    limb. The dashed box denotes the FoV of the VM used for
    extrapolation (see \Fig~2). The inserted panels are observations
    of the filament channels in AIA-304 and the H$\alpha$ filament at
    different times.}
  \label{fig1}
\end{figure*}

\begin{figure*}[htbp]
  \centering
  \includegraphics[width=0.8\textwidth]{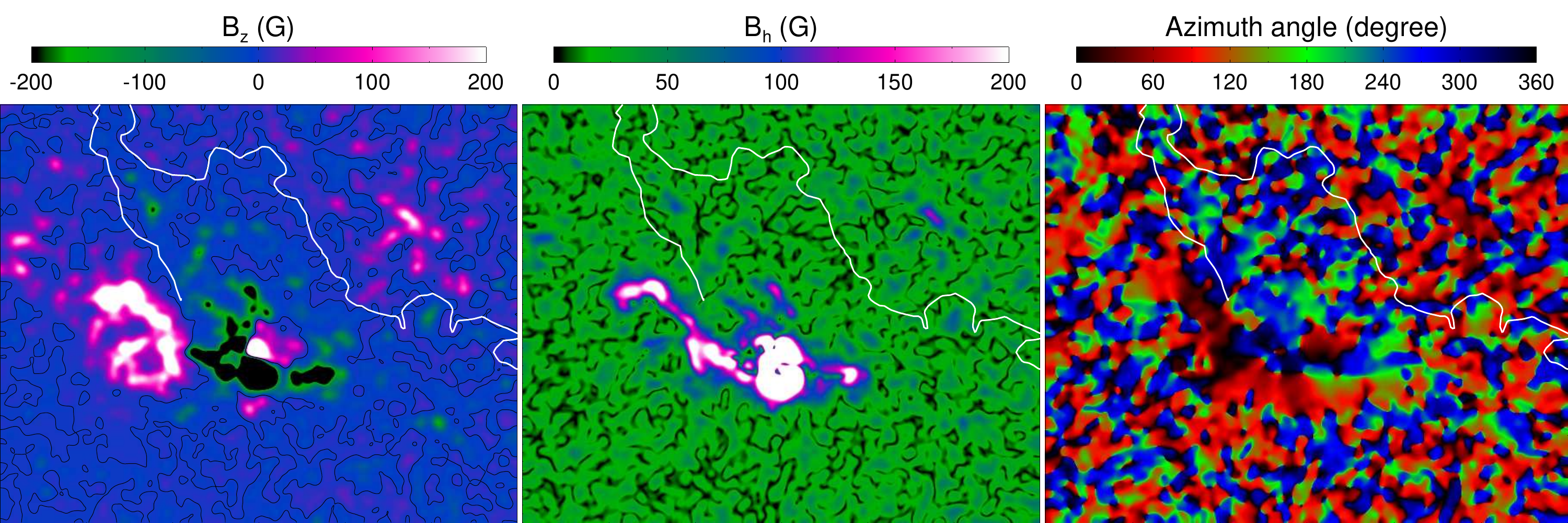}
  \caption{HMI VM at 22:00~UT on September 6. Its FoV is outlined by the dashed box in \Fig~1. For left to right are vertical field strength, horizontal field strength, and azimuthal angle, respectively. The map shown here is Gaussian-smoothed from the original data with FWHM of 10~arcsec. The photospheric PILs are shown by the black contours in the left panel. The white thick lines overlaid on the maps are the PIL derived from the potential field extrapolated to a height of 10~arcsec, and only the part along the filament channel is shown.}
  \label{fig2}
\end{figure*}

\section{Observations}
\label{sec:obse}

A sigmoidal filament was observed to the northwest of AR~11283 during
its passage of the solar disk. \Fig~1 shows the observations in BBSO
H$\alpha$ and SDO/AIA-304. The filament first appeared at the east
limb on 2011 August 30. It was possibly formed before rotating into
the visible disk. The main body of the filament seen in H$\alpha$
shows a slightly inverse-S shape with a length of about 200~Mm.  The
filament channel is more extended, as observed by AIA, roughly forming
a horizontally-lying, full inverse S-shape.
Its west part is much more evident than the east part which appears
rather fragmented. The filament was stable for days and partially
erupted on early September 8 when approaching the west limb. Before
its eruption, although the H$\alpha$ filament evolved, the large-scale
shape of the filament channel almost showed no clear change. It
suggests that the basic underlying magnetic structure is very stable.

The H$\alpha$ observations show that
there are barbs along the filament mainbody, and most of them
are right-bearing. This indicates the filament's
chirality as dextral,
meaning that the axial magnetic field in the filament
points to the right when viewed from the side with positive polarity
in the photosphere \citep{Pevtsov2003}. It can be seen from \Fig~1 that the
photospheric field in the southeast of the filament is predominately
negative and in the northwest predominately positive. Hence according
to the chirality, the axial field in the filament mainbody would point to
the northeast. We will discuss more details of the filament structure
with the help of the extrapolated coronal field.


\begin{figure*}[htbp]
  \centering
  \includegraphics[width=0.8\textwidth]{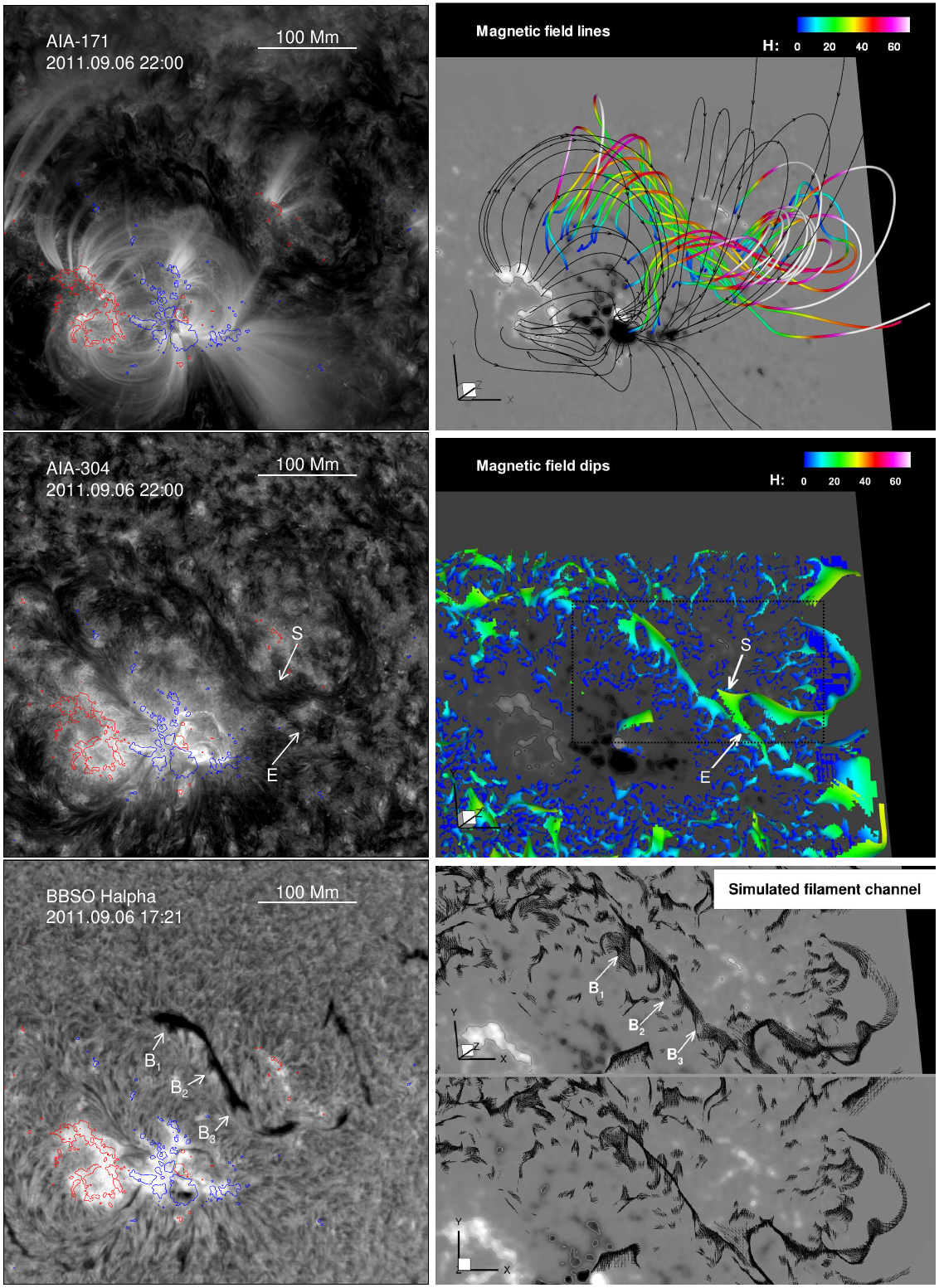}
  \caption{Left: observations of the filament channel by SDO/AIA and
    BBSO H$\alpha$. Contours overlaid are LoS photospheric magnetic
    field (red for 200~G and blue for $-200$~G). Right: model results
    of the coronal field. The magnetogram of $B_{z}$ is shown as the
    gray background. In the right top and middle panels, the color
    represents the height from the bottom (unit in Mm). {\it see the
      text for details.}}
  \label{fig3}
\end{figure*}
\begin{figure*}[htbp]
  \centering
  \includegraphics[width=0.8\textwidth]{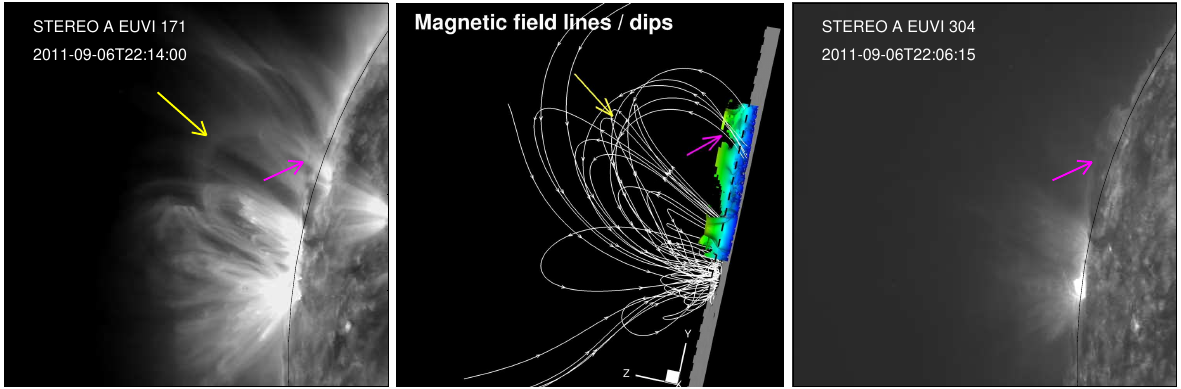}
  \caption{Left and right: STEREO-A observations of the prominence in
    EUVI-171 and EUVI-304, respectively. The black arcs are the solar
    limb. The pink arrow marks the prominence, and the yellow arrow
    marks the overlying arcades. Middle: the coronal field lines model
    the EUVI-171 loops and magnetic field dips model the prominence,
    with the solar limb denoted by the dashed black line. Note that
    only the dips of the FR main body, i.e., those
    within the dashed box in middle right panel of
    \Fig~3, are shown here.}
  \label{fig4}
\end{figure*}

\begin{figure*}[htbp]
  \centering
  \includegraphics[width=0.8\textwidth]{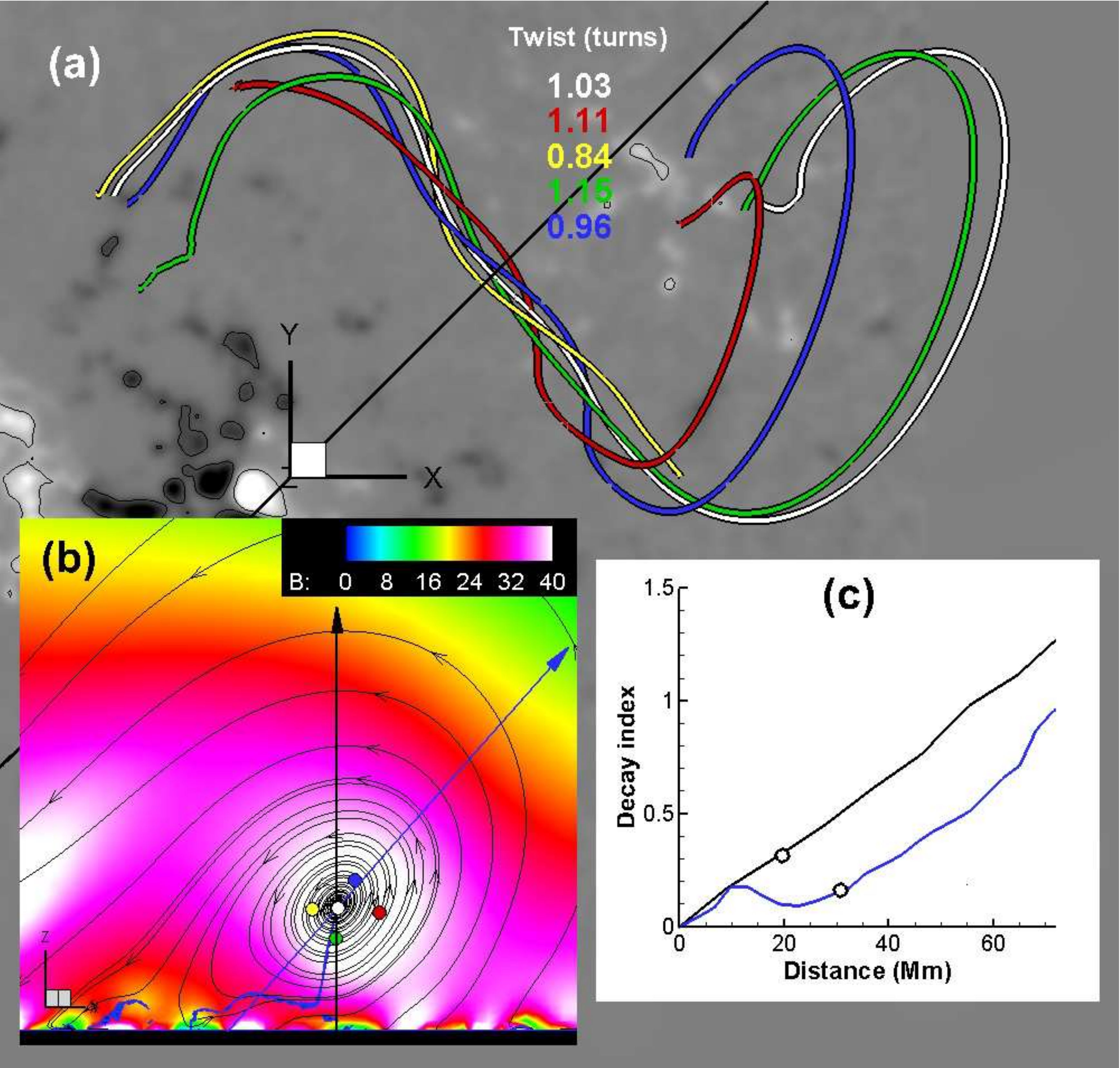}
  \caption{(a) Four sampled field lines (colored by blue, red, green and
    yellow, respectively) around the FR axis (white). The colored numbers
    denote the twist degrees of these field lines.
    (b) Central cross section of the FR, which is vertically sliced
    along the black line shown in panel (a).
    The streamlines show the 2D field-line tracing on the slice, which
    forms helical lines centered at the axis of the FR. The background
    shows the magnetic field strength (unit in G). The small colored
    circles denotes the locations of the field lines shown in panel
    (a). Note that the blue, thick curved lines consist of the magnetic
    dips. 
    The two directed lines through the helix center (i.e., the FR
    axis) represent the paths along which we calculate the decay
    index. Distributions of the decay index along these two paths
    (starting from the bottom) are displayed in panel (c), and the
    locations of the rope axis are marked by the circles on the lines
    of the decay index.}
  \label{fig5}
\end{figure*}

\section{Results}

Here we only show NLFFF extrapolation at 22:00~UT on September 6. This
moment interests us because there is a sigmoid eruption from AR~11283,
which has been studied formerly in \citep{Jiang2013MHD,
  Jiang2014formation}, and obtaining the large-scale coronal field
around this moment will improve our understanding of the sigmoid
eruption. \Fig~2 shows the VM. It has a field of view (FoV) of
600$\times$480~arcsec$^2$, which is large enough to include almost the
whole filament channel (except that the west end of the channel is
outside of the FoV). The magnetogram as shown has been
Gaussian-smoothed with an FWHM of 10~arcsec since the original data is
much more noisy. Even though, we find that photospheric PILs are still
too fragmented to identify where the filament is located above.  To
put a reasonable PIL that follows the filament channel, we overlay on
the maps the potential-field PIL computed at a small height
(10~arcsec) above the photosphere, which is much more extended than
the photospheric PIL because the small polarities decay fast with
height. As can be seen along this PIL that represents the path of the
EUV filament channel, the transverse field is rather weak ($\le
100$~G), and its directions show a rather random pattern without
systematic magnetic shear regarding the length scale of the filament.


The extrapolation is carried out using our CESE--MHD--NLFFF code in
exactly the same way as done by \citet{Jiang2013NLFFF}, without
any parameter optimized for the present modeling. The code is based on
MHD relaxation method and implemented by an advanced
conservation-element/solution-element (CESE) spacetime
scheme on an 
adaptive grid with parallel computation \citep{Jiang2010}.
Before inputting to the
extrapolation code, the original VM is preprocessed to remove the
photospheric Lorentz force using a new preprocessing method \citep{Jiang2014pre}.
We compute the field in a Cartesian box of $768\times 640\times 384$~arcsec$^3$ (with
resolution of 1~arcsec), which includes a peripheral region around the
VM\footnote{In principle it would be better to compute the field in
  the spherical geometry for such a large FoV \citep{Jiang2012apj1}, so the extrapolated
  field can be accurately co-aligned with observations.}.

\subsection{Extrapolation result and comparison with observations}

\Fig~3 compares the extrapolation result with AIA and H$\alpha$
observations. The top-right panel shows the magnetic field lines. The
angle of view is co-aligned approximately with the SDO. The black
field lines are plotted to represent the coronal loops observed in
AIA-171. Along the filament channel there is indeed a FR, which is
shown by the colored rod-like lines, with the color denoting the
height from the photosphere as indicated by the colorbar. As can be
seen, the middle of the FR is low-lying at about $20\sim 40$~Mm, while
its two ends reach highly and are close to arcades. There is some flux
of the rope passing through the west boundary, possibly because the
FoV of the magnetogram does not fully include the entire FR
system. Overall the rope presents an inverse S-shape roughly matching
the filament channel, and its chirality is correctly reproduced. It is
interesting to find that some field lines from the AR's sunspot turn
into the FR.  Overlying the FR is a group of potential-like field
lines, and only the two ends of these field lines are visible like
bright rays in the AIA-171 image, because the emission drops off
rapidly with height. This overlying closed arcade is expected to play
a key role in stabilizing the FR.

The middle-right panel of \Fig~3 shows the magnetic dips, i.e., the
locations where the field lines are locally horizontal ($B_z = 0$) and
curved upwardly ($\vec B \cdot \grad B_z \ge 0$). The dips are also
pseudo-colored by the height. In the last two right panels of \Fig~3,
we plot the horizontal field vectors (using an uniform length of 1~Mm)
at all the dips to simulate the observed filament channel
\citep{Ballegooijen2004}. Evidently, there is a long extended dip
reaching above 30~Mm, reproducing very well the main body of the
quiescent filament (i.e., the middle section of the filament
channel). Moreover, a sigmoidal channel, following the observed
inverse S-shaped filament channel, can also be observed from the dips,
although the east end is very fragmented as in the case of
observation. The model does not reproduce well the west end of the
filament channel because it is near the boundary of the magnetogram.

In the middle panels as marked by the arrow $S$, it appears that the
filament spine is interrupted by a north deviation of the west part
from the main body, resulting in an elongated barb in the south side
(marked by the arrow $E$). Our model recovers this offset and the
elongated barb, and they are caused by a local intrusion in the PIL by
a group of strong positive polarities ($>100$~G, just at the site of
the interruption, see \Fig~1), which locally alters the path of the
FR. In addition, the right-bearing barbs along the filament mainbody,
as marked by the arrows of $B_1$, $B_2$ and $B_3$, are reproduced by
the model. As shown by the model result viewed from above (Fig~3,
bottom right panel), there are also several `barbs' at the north side,
which are not seen in the observation because they were hidden behind
the filament spine.

By inspecting the dips shown in the middle-right panel, we find that
there are a large number of localized fragmentation of dips (in deep
blue color) that is very close to and on the photosphere (i.e., bald
patches). These dips result from the highly intermittent distribution
of the photospheric field \citep{Aulanier2002, Dudik2008}. The number of the localized dips decreases
very fast with height since the related small-scale fields decay fast
with height. Around the AR, the model produces many dips of rather
short extensions, and these dips may support the small-scale filaments
as observed around the AR both in AIA and H$\alpha$.
We note that high dips near the side boundaries should not be
taken seriously because the magnetic field on those boundaries are
fixed, and during NLFFF computing, field lines near the boundaries are
squeezed to produce these artificial dips.

The filament was also observed by STEREO-A at the limb as a low-lying
prominence. \Fig~4 compares the STEREO observations with the model
result shown also in a limb view. The field lines compare the EUV
bright loops and the dips represent the prominence. As viewed by
STEREO-A, a closed arcade (marked by a yellow arrow) overlying the
prominence is clearly observed in the EUVI-171. This arcade
corresponds to the potential-like field lines, two ends of which are
visible in AIA-171. The prominence reaches highest (about 30~Mm) in
the north, then smoothly descends along its path to the south, and its
spine turns into the solar disk. The modeled magnetic dips reproduce
correctly the height profile of the observed prominence, except that
near the south end the simulated prominence appears much higher than
the observed one. We attribute such disagreement to several reasons:
first, the model does not construct properly the west end of the
filament channel because it is beyond the FoV of the magnetogram;
second, the model is constructed in Cartesian box, as a result the
curvature of the long filament channel is not well characterized;
last, the magnetic dips do not necessarily represent the prominence if
they are not filled with the cold prominence plasma.

\subsection{Stability of the FR}

It is known that a coronal FR confined by overlying potential arcade
is subject to two kinds of instabilities, i.e., the kink instability
(KI) and torus instability (TI). The KI states that a FR will deform
helically if the twist degree, which is measured by the number of
windings of the field lines around the rope axis, exceeds a critical
value \citep[about 1.5--2,][]{Torok2005}. The TI occurs when the
outward expansion of a FR can no longer be confined by the overlying
field if it decays faster than an unstable threshold, which is
characterized by a decay index of the external field with an unstable
threshold of $\sim 1.5$ \citep{Kliem2006}.

In \Fig~5 (a), we show several field lines around the FR's axis with
different colors. These field lines pass through the central cross
section of the FR, as shown in \Fig~5 (b), and the rope axis is
located at the center of the helical shapes formed by the poloidal
flux of the rope. Note that in \Fig~5 (b) the background shows the
magnetic field strength. It can be clearly noted that the field
strength is enhanced in the rope region because of the strong axial
flux of the rope. The cross section of the field dips (the thick blue
curved line) starts from the FR axis and reaches down to the photosphere.
The FR is tilted to the right due to the non-symmetry of the flux
distribution. Consequently, the field dips are aligned not vertically,
but with a small deviation of $\sim 10^{\circ}$ from the vertical. Due
to a localized parasitic polarity intrusion at the photosphere PIL,
the dips reaching the bottom are strongly deflected to almost in a
horizontal way, producing naturally a lateral foot, i.e., barb of
the filament \citep{Aulanierfilament1}.

The twist degree of the FR is computed according to \citep{Inoue2011},
$T_n=\int \alpha dl/(4\pi)$, where $T_n$ corresponds to the number of turns
of the magnetic field line, $\alpha$ is the force-free parameter and the integral
$\int dl$ is taken along the magnetic field line. 
The result shows that the field lines make about one turn (see \Fig~5 (a)),
clearly less than the threshold for KI. To test if the TI can occur,
we compute the decay index of the potential field in the central cross
section. The decay index is calculated along two different lines
through the rope axis, a vertical and an inclined one that is
approximately along the inclination of the overlaying arcade. Note
that only the component of the potential field perpendicular to the
directions is used \citep{Jiang2014formation}. \Fig~5 (c) shows the
result, and the location of the rope axis is marked by circles on the
profile lines of the decay index. Clearly the FR is far below the
height with TI threshold, i.e., 1.5. So the FR is very strongly
confined by its overlying flux, and a slow evolving of the
photospheric field is not likely to trigger the filament eruption on
September 8. We suggest that it is instructive to study the eruption
in the context of the interaction between the AR eruption and the
quiescent filament.

\section{Conclusions}
\label{sec:colu}

This Letter reports an NLFFF extrapolation of solar coronal field that
holds a large-scale filament from photospheric VM, which provides
concrete evidence for filaments being supported by magnetic
FR. Although the presence of a large-scale FR can hardly be predicted
from the noisy VM, our CESE-MHD-NLFFF extrapolation code is able to
overcome this difficulty
to extract the key information of the coronal field from the
magnetogram. We have also examined the robustness of the extrapolation
by using VMs at two other different times nearby (not shown here),
which also reproduce the similar FR structure. A detailed comparison
with multiple observations, including those of stereoscopic
viewpoints, demonstrates that the filament structure is well
reproduced by the extrapolation. The FR that supports the filament is
very stable because it is weakly twisted and strongly confined by the
overlying closed arcades. Its eruption is likely triggered by a nearby
AR eruption, which awaits further investigation.




\acknowledgments

This work is supported by NSF-AGS1153323, AGS1062050 and in addition
C.W.J and X.S.F are also supported by the 973 program under grant
2012CB825601, the Chinese Academy of Sciences (KZZD-EW-01-4), the
National Natural Science Foundation of China (41204126, 41231068,
41274192, 41031066, and 41374176), and the Specialized Research Fund
for State Key Laboratories. Data from observations
are courtesy of NASA {SDO}/AIA and the HMI science teams.


\end{CJK*}
\end{document}